\documentstyle[prb,twocolumn,epsfig,floats,aps]{revtex}
\begin{document}
\draft

\twocolumn[\hsize\textwidth\columnwidth\hsize\csname @twocolumnfalse\endcsname

\title{Classical percolation transition \\
in the diluted two-dimensional $S=1/2$ Heisenberg antiferromagnet}

\author{Anders W. Sandvik}
\address{Department of Physics, {\AA}bo Akademi University, Porthansgatan 3,
FIN-20500, Turku, Finland}

\date{\today}

\maketitle

\begin{abstract}
The two-dimensional antiferromagnetic $S=1/2$ Heisenberg model with random 
site dilution is studied using quantum Monte Carlo simulations. Ground state 
properties of the largest connected cluster on $L \times L$ lattices, with $L$ 
up to $64$, are calculated at the classical percolation threshold. In addition,
clusters with a fixed number $N_c$ of spins on an infinite lattice at the
percolation density are studied for $N_c$ up to $1024$. The disorder averaged 
sublattice magnetization per spin extrapolates to the same non-zero 
infinite-size value for both types of clusters. Hence, the percolating 
clusters, which are fractal with dimensionality $d=91/48$, have 
antiferromagnetic long-range order. This implies that the order-disorder 
transition driven by site dilution occurs exactly at the percolation 
threshold and that the exponents are classical. The same conclusion is reached
for the bond-diluted system. The full sublattice magnetization versus site
dilution curve is obtained in terms of a decomposition into a classical 
geometrical factor and a factor containing all the effects of quantum 
fluctuations. The spin stiffness is shown to obey the same scaling as
the conductivity of a random resistor network.

\end{abstract}


\vskip2mm]

\section{Introduction}

The two-dimensional (2D) Heisenberg antiferromagnet on a square lattice can be
driven through a quantum phase transition \cite{chn,sachdev} by, e.g., 
introducing frustrating interactions \cite{frust} or by dimerizing the 
lattice.\cite{dimer} It has also been believed that a non-trivial (quantum) 
phase transition could be achieved by diluting the system, i.e., by randomly 
removing either sites \cite{manousakis,behre,yasuda,castro} or bonds.
\cite{wan,awsmv} The site dilution problem is of direct relevance in the 
context of antiferromagnetic layered cuprates doped with nonmagnetic 
impurities.\cite{doping,chernyshev,greven} Diluted Heisenberg models are also 
of more general interest, as systems in which the combined effects of disorder 
and quantum fluctuations can be studied with a variety of analytical and 
numerical methods. The single impurity problem has been studied extensively
and is now rather well understood.\cite{singleimp} Systems with a finite 
concentration of impurities are much more difficult to treat, both 
analytically and numerically. The location and nature of the phase
transition driven by dilution is therefore still controversial.

An early quantum Monte Carlo (QMC) study of the temperature dependence of the 
correlation length gave a bound $p_c > 0.2$ for the critical fraction 
of removed sites above which the long-range order vanishes in the 2D 
Heisenberg model.\cite{manousakis} QMC calculations in the ground state 
indicated $p_c \approx 0.35$.\cite{behre} Various analytical treatments have 
given results for $p_c$ ranging from $0.07$ to $0.30$.\cite{yasuda,castro} 
These estimates for the critical hole concentration are below the classical 
percolation threshold $p^* \approx 0.407$,\cite{stauffer,percdens} and hence 
the phase transition would be caused by quantum fluctuations. A critical
hole density much smaller than the percolation density was also found in
the bond diluted Heisenberg model.\cite{wan,awsmv} 

An unusual type of quantum phase transition in the site diluted system was 
recently claimed by Kato {\it et al}.\cite{kato} They 
carried out QMC simulations of larger lattices at lower temperatures than in 
previous works and found evidence of the critical dilution coinciding with 
the classical percolation point; $p_c = p^*$. In spite of this, they argued 
that the transition is a non-trivial quantum phase transition, which would be 
a consequence of the fractal clusters at $p^*$ being quantum critical (i.e., 
with algebraically decaying spin-spin correlation function). This leads to 
non-classical critical exponents, which furthermore were found to be 
non-universal, dependent on the spin $S$ of the magnetic sites 
(approaching the classical values when $S\to\infty$). Although such behavior 
violates the standard notions of universality, it cannot be completely 
excluded for random systems.\cite{suzuki} However, in another recent study 
the spin correlations of the percolating 2D Heisenberg model with $S=1/2$ 
were analyzed in greater detail.\cite{awscomment} It was confirmed that 
$p_c \equiv p^*$, but, in conflict with the quantum criticality scenario,
\cite{kato,todo,yasuda2} strong evidence was presented of a transition
driven solely by percolation. The exponents should then be identical to 
those of classical percolation for all $S$. 

This paper presents details of the QMC studies highlighted in
Ref.~\onlinecite{awscomment} and introduces further evidence that the
order-disorder transition in the diluted 2D Heisenberg model indeed occurs 
exactly at $p^*$ and is classical. The stochastic series expansion (SSE) QMC 
method \cite{sse1,sse2,sse3} is used to study the ground state of both site
and bond diluted systems at their respective percolation points. Site 
diluted systems are also studied for the whole range of hole concentrations 
$p < p^*$. Particular emphasis is put on the importance of carefully 
controlling potential sources of systematic errors in the simulations. 
In studies of disordered systems these issues are much more serious than 
for clean systems, because of the necessity to carry out a large number of 
relatively short simulations for different samples (in order to obtain 
accurate disorder averages). Procedures developed to accelerate the 
equilibration, and to detect possible remaining effects of insufficient 
equilibration and finite temperature, are discussed here and constitute 
an important part of the paper. 

The main physics questions addressed and results obtained are summarized as 
follows. At the percolation point, the infinite clusters on a 2D lattice have 
a fractal dimensionality, $d=91/48$.\cite{stauffer} An antiferromagnet at 
this special point could in principle be either classically critical (if there
is long-range order on the fractal clusters), quantum 
critical (with power-law decaying spin-spin correlation function on the 
clusters), or quantum disordered (with exponentially decaying 
correlations on the clusters). In the last of these cases,
the phase transition would occur at a dilution fraction
less than the percolation density, whereas it coincides with the percolation 
point in the other two cases. In order to determine which of the three 
qualitatively different ground states is realized in the percolating cluster
of the standard Heisenberg model, the sublattice magnetization is calculated 
for the largest cluster on $L \times L$ lattices at the percolation density, 
with $L$ up to $64$. In addition, clusters of fixed size $N_c$ without 
boundary imposed shape constraints (i.e., on an infinite 2D lattice) are 
studied for $N_c$ up to $1024$. The sublattice magnetization is averaged over 
several thousand samples and extrapolated to infinite size. The same non-zero 
value is obtained for both types of clusters, showing consistently that they 
are long-range ordered. Self-averaging is demonstrated by studying 
sample-to-sample distributions of the sublattice magnetization. The existence 
of long-range order on the percolating clusters implies that the 
order-disorder transition driven by dilution occurs exactly at the percolation
threshold and that the critical exponents are classical. The same qualitative
behavior is found for site and bond dilution, but the long-range order is
substantially weaker in the bond diluted system.

In order to reliably calculate the experimentally interesting sublattice 
magnetization $M$ as a function of the site dilution fraction $p$ for all 
$0 \le p < p^*$, a decomposition of $M(p)$ into a classical and a quantum 
mechanical factor is used. The classical factor, which contains the singular 
behavior at $p=p^*$, can be easily evaluated by classical Monte Carlo. The 
critical exponent governing its asymptotic $p \to p^*$ form is known exactly. 
\cite{stauffer} The quantum mechanical factor is calculated using QMC 
simulations of the largest cluster on $L\times L$ lattices. It is only 
weakly dependent on the dilution. The whole $M(p)$ curve is determined to 
an accuracy of a few percent.

The spin stiffness is also calculated. Based on known results for the
classical Heisenberg model \cite{harris1} and the long-range order found 
here in the percolating clusters, it is argued that the stiffness should 
obey the same scaling as the conductivity of a random resistor network at 
and close to the percolation threshold. The numerical results are fully 
consistent with the known conductivity exponent.

The outline of the rest of the paper is the following: In Sec.~II the various 
types of diluted Heisenberg lattices are defined, and the application of the 
SSE simulation algorithm to these systems is discussed. The procedures 
developed for controlling potential systematic errors arising from 
insufficient equilibration and finite temperature are also introduced here. 
Simulation data illustrating the convergence criteria are presented in 
Sec.~III. In Sec.~IV the sublattice magnetization of percolating clusters is 
studied, both for site and bond diluted systems. In Sec.~V the full 
sublattice magnetization versus site-dilution curve is calculated. Results
for the spin stiffness are presented in Sec.~VI. The paper concludes 
with a summary and discussion in Sec.~VII.

\begin{figure}
\centering
\epsfxsize=6.5cm
\leavevmode
\epsffile{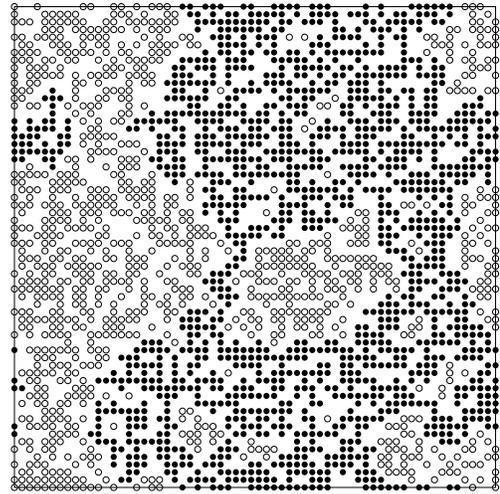}
\vskip2mm
\caption{A $64\times 64$ lattice randomly diluted at $p=p^*$. The solid
circles indicate magnetic sites belonging to the largest connected cluster
(note that periodic boundary conditions are applied). The other magnetic 
sites are shown as open circles.}
\label{fig:lconf}
\end{figure}

\section{Models and Methods}

The antiferromagnetic $S=1/2$ Heisenberg model on several types of random 
2D lattices will be considered. In all cases, the Hamiltonian can be written 
in the form
\begin{equation}
H = J\sum_{b=1}^{N_b} {\bf S}_{i(b)} \cdot {\bf S}_{j(b)},
\label{ham}
\end{equation}
where $b$ is a bond index corresponding to two interacting nearest-neighbor 
spins $i(b),j(b)$ and $N_b$ is the total number of bonds on the lattice. On 
a site diluted lattice a fraction $p$ of the sites are empty (holes) and the 
rest are occupied by spins. Bonds exist between all occupied nearest neighbor 
sites. On a bond diluted lattice all sites are occupied and nearest neighbors 
interact with a probability $p$. Note that a diluted lattice typically 
contains isolated (free) spins that are not interacting with any other spins. 
They have to be specified in addition to the list of bonds $\{i(b),j(b)\}$ 
in the Hamiltonian (\ref{ham}).

\begin{figure}
\centering
\epsfxsize=8.3cm
\leavevmode
\epsffile{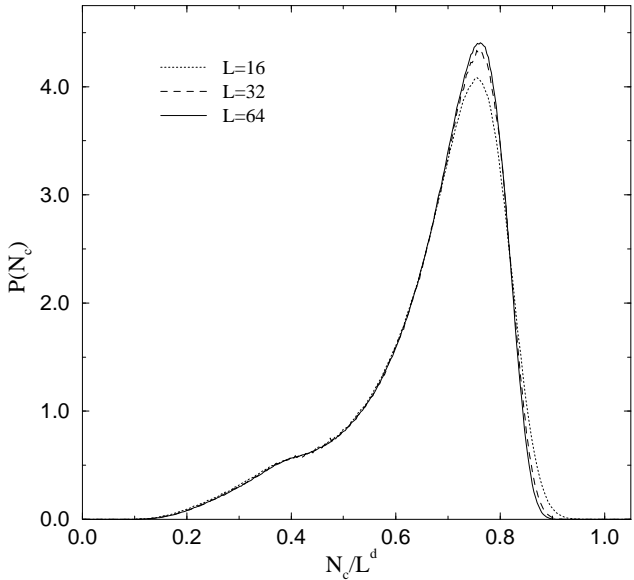}
\vskip1mm
\caption{Distribution of the size of the largest cluster on periodic 
$L\times L$ lattices for $L=16,32$, and $64$. The probability $P(N_c)$ of 
cluster size $N_c$ is graphed versus $N_c/L^d$, showing scaling with the 
fractal dimension $d=91/48$. Note the structure at $N_c/L^d\approx 0.38$, 
which corresponds to lattices where instead of one dominant large cluster 
there are two of approximately half the size.}
\label{fig:nc}
\end{figure}

\subsection{Diluted lattices}

For lattices with $N=L\times L$ sites and periodic boundary conditions, random
magnetic configurations (samples) are generated by filling each site with 
probability $1-p$. The actual number of magnetic sites is hence not fixed, but
the fluctuations in the density decrease as $1/L$. The percolation density 
$p=p^*$ is of special relevance. According to the most recent simulation,
\cite{percdens} $p^*=0.40725379(13)$. Here the value $p^*=0.4072538$ will be 
used. The largest cluster of connected magnetic sites is of particular 
interest and its properties will be studied separately from those of the full
lattice. The number of spins belonging to the largest cluster is denoted by 
$N_c$. At $p=p^*$, in the limit $L\to \infty$, this cluster is fractal, with 
the fractal dimension $d$ known rigorously to be $d=91/48$.\cite{stauffer} For
large $L$ the average $\langle N_c \rangle \sim L^d$, and $N_c$ is therefore 
typically considerably smaller than the total number of spins on the lattice. 
One can therefore reach larger cluster sizes in the QMC simulations by 
removing the spins that do not belong to the largest cluster. This will be 
done here in order to study the clusters for $L$ as large as $64$. An 
example of a a diluted lattice and its largest cluster is shown in 
Fig.~\ref{fig:lconf}.

\begin{figure}
\centering
\epsfxsize=6.5cm
\leavevmode
\epsffile{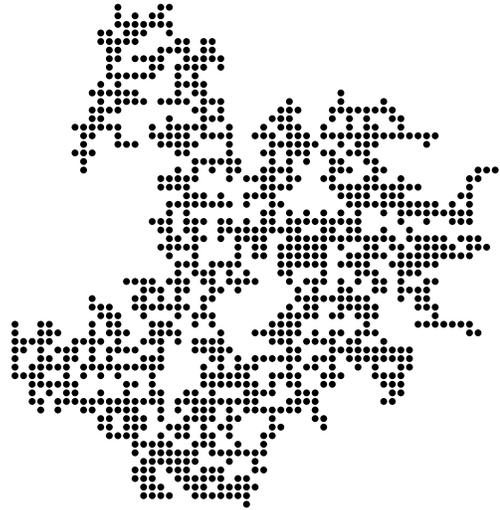}
\vskip2mm
\caption{A cluster with $N_c=1024$ sites constructed on an
infinite 2D lattice at the percolation density.}
\label{fig:nconf}
\end{figure}

The largest cluster on a lattice at $p=p^*$ exhibits strong size 
fluctuations, as shown in Fig.~\ref{fig:nc}. As an alternative to approaching 
the infinite fractal lattice as a function of $L$ with fluctuating $N_c$, 
clusters with fixed $N_c$ and shapes not restricted by lattice boundaries will
also be studied. Such clusters are constructed starting from an infinite 2D 
lattice with only a single filled site. The four neighbors of this site are 
filled at random with probability $1-p^*$. In the next step the neighbors of 
those sites that were filled are in turn filled with probability $1-p^*$, 
taking into account that sites that were previously visited should not be 
visited again. This procedure is repeated until no new sites can be filled 
that are connected to the cluster, i.e., the nearest neighbors of all sites 
in the cluster have already been visited. If the cluster is completed before 
it reaches the desired size $N_c$, or if the size exceeds $N_c$, the cluster 
building is restarted. The process is repeated until a cluster is completed 
exactly at the size $N_c$. This method of constructing fixed-size clusters 
becomes very time consuming for large $N_c$, but it works well for sizes 
$N_c \le 1024$ considered here. An example of this type of cluster is shown 
in Fig.~\ref{fig:nconf}.

In the case of bond dilution, the percolation point is exactly 
$p^*=1/2$.\cite{stauffer} For $L \times L$ lattices this probability can be 
realized for any $L$ and therefore random lattices with exactly half of the 
bonds removed will be considered in calculations at the percolation 
threshold. 

\subsection{Quantum Monte Carlo algorithm}

The SSE approach to QMC simulation of lattice models\cite{sse1} has been 
discussed in detail in previous papers. Its application to the Heisenberg 
model is discussed in, e.g., Refs.~\onlinecite{sse2,sse3,sign}. Its 
effectiveness for various ordered and disordered systems has recently been 
documented by several groups.\cite{sseappl,wessel,clay,dorneich} Here only a 
very brief summary will be given, in order to facilitate the subsequent 
discussion of procedures developed for efficient equilibration and ground 
state convergence for disordered systems.

In order to apply the SSE method to the Heisenberg model, the Hamiltonian
(\ref{ham}), with $J=1$ hereafter, is first written as
\begin{equation}
H = -\sum_{b=1}^{N_b} [H_{1,b} - H_{2,b}],
\label{ham2}
\end{equation}
where the pair interaction has been divided into terms 
\begin{eqnarray}
H_{1,b} & = & \hbox{$1\over 4$} - S^z_{i(b)}S^z_{j(b)}, \label{h1} \\
H_{2,b} & = & \hbox{$1\over 2$}[S^+_{i(b)}S^-_{j(b)} + S^-_{i(b)}S^+_{j(b)}],
\label{h2}
\end{eqnarray}
which are diagonal and off-diagonal, respectively, in the basis
$\{| \alpha \rangle \} =  \{ | S^z_1,S^z_2,\ldots ,S^z_N \rangle \}$
used in the simulations.
A constant has been added to the diagonal part, and as a result all 
non-vanishing matrix elements equal $1/2$ and correspond to operations 
on anti-parallel spins. 

The SSE algorithm is based on importance sampling 
of the terms of the partition function $Z={\rm Tr} \{ {\rm e}^{-\beta H} \}$
written in a truncated Taylor expansion form:
\begin{equation}
Z=\sum\limits_\alpha \sum\limits_{S_M} {\beta^n (M-n)! \over M!}
\langle\alpha | \prod_{i=1}^M
H_{a_i,b_i} |\alpha\rangle .
\label{zl}
\end{equation}
The summation symbol $S_M$ refers to a sequence of $M$ operator-index pairs, 
\begin{equation}
S_M = [a_1,b_1],[a_2,b_2],\ldots ,[a_M,b_M], 
\end{equation}
where $a_i \in \{1,2\}$, $b_i \in \{1,\ldots N_b \}$, corresponding to the
operators $H_{a_i,b_i}$ in (\ref{h1}) and (\ref{h2}), or $[a_i,b_i]=[0,0]$, 
corresponding to an identity operator $H_{0,0} \equiv I$. This new operator 
has been introduced in order for the summation over all $S_M$ in (\ref{zl}) 
to imply summation of the Taylor expansion of ${\rm e}^{-\beta H}$ up to 
order $M$. The order of a given term corresponds to the number of non-$[0,0]$ 
elements in $S_M$, which is denoted by $n$ in (\ref{zl}). It has been assumed 
that the lattice is bipartite. All the signs arising from the off-diagonal 
operators $H_{2,b}$ in (\ref{ham2}) then cancel in the non-vanishing terms of 
(\ref{zl}) and the expansion is hence positive-definite. The cut-off $M$ can 
be easily adjusted so that $n$ never reaches $M$ during the simulation. The 
truncation then does not constitute an approximation, and SSE simulation 
results are thus exact to within statistical errors. As will be explained 
further below, $M$ has to be chosen proportional to $N\beta$. 

For the sampling of the terms $(\alpha,S_M)$ an efficient algorithm with
three basic updates has been developed. The first update involves only
diagonal operators. The sequence $S_M$ 
is scanned from $i=1$ to $M$, and for each element $[a_i,b_i]$ with $a_i=0$ 
or $a_i=1$ a substitution $[0,0] \leftrightarrow [1,b_i]$ is attempted. 
The Metropolis acceptance probability can be easily calculated from 
Eq.~(\ref{zl}), taking into account also that an update in the $\rightarrow$ 
direction is allowed only if the spins at the tentative bond $b_i$ are 
anti-parallel after operation with the previous $i-1$ operators. An accepted
single-operator update changes the expansion power $n$ in (\ref{zl}) by 
$\pm 1$. 

The second update is a more complicated cluster-type update which operates at 
fixed $n$ and simultaneously changes the operator-type index of several 
elements $\{i \}$. The set $\{i \}$ forms an ``operator loop'', the size of 
which can be very large. For each $i$ the substitution $[1,b_i] 
\leftrightarrow [2,b_i]$ can be carried out without changing the 
configuration weight. The whole sequence $S_M$ can be uniquely decomposed 
into a number of operator-loops, which can be updated independently of each 
other with probability $1/2$. Details of this operator-loop update are
discussed in Ref.~\onlinecite{sse3}.

Spins in the state $| \alpha\rangle$ that are not acted upon by any operator 
in $S_M$ are flipped with probability $1/2$. Apart from isolated spins on a 
diluted lattice, such free spins appear frequently only at high temperatures. 

With the three updates described above --- single-operator (or diagonal), 
operator-loop, and spin flip --- the SSE method is completely grand canonical,
i.e., all magnetization and winding number sectors are sampled. In systems 
with no isolated spins, the spin flip is strictly not needed, but it is still 
useful at high temperatures. 

The simulation is started with an arbitrary state $|\alpha\rangle$ and a short
index sequence containing only $[0,0]$ elements (any $M$ will do --- in the 
work discussed here $M=N_b/4$ was typically used). $M$ is adjusted during the 
equilibration of the simulation, so that it always exceeds the maximum 
$n$ reached (by, e.g., 20$\%$), and is thereafter kept constant. 
A Monte Carlo step (MC step) consists of a full sweep of single-operator 
updates followed by construction and updates of all operator loops. After 
this, free spins in the state $| \alpha\rangle$ are flipped with probability 
$1/2$. Further details of the sampling procedures have been described in 
Refs.~\onlinecite{sse3} and \onlinecite{sign}. 

In the computer, an operator 
$[a,b]$ can be represented by a single 4-byte integer. In addition, 
in the cluster update four integers are needed to store each operator
element in $S_M$ with their pointers to other elements in the list.\cite{sse3} 
The total memory requirement is thus $20\times M$ bytes,\cite{memorynote} 
plus a few arrays the sizes of which scale linearly with the system size 
$N$. The number of operations needed for carrying out one MC step scales 
as $M$, i.e., is proportional to $N\beta$.

Observables are typically measured after every MC step (it is often practical 
to do the calculations in combination with the single-operator update).
Estimators for various expectation values of interest in the context of the 
Heisenberg model have been discussed in Ref.~\onlinecite{sse2}. In the
present work, the most important quantity is the staggered structure factor, 
defined on the whole $L\times L$ lattice as (for a given disorder realization,
with $S^z = 0$ on the non-magnetic sites)
\begin{equation}
S(\pi,\pi) = {1\over N}\left\langle\left (
\sum\limits_{i=1}^{N} (-1)^{x_i + y_i} S^z_i \right )^2 \right\rangle ,
\label{spi}
\end{equation}
and on the largest cluster $C$  (or the single cluster on the infinite 
lattice)
\begin{equation}
S_{c}(\pi,\pi) = {1\over N_c} \left\langle \left (
\sum\limits_{i\in C} (-1)^{x_i + y_i} S^z_i \right )^2 \right\rangle .
\label{spic}
\end{equation}
Disorder averaged sublattice magnetizations are defined in terms of the 
structure factors according to
\begin{eqnarray}
\langle m^2 \rangle & = & \langle 3S(\pi,\pi)/N \rangle, \label{subm} \\
\langle m_c^2 \rangle & = & \langle 3S_c(\pi,\pi)/N_c \rangle \label{submc},
\end{eqnarray}
where, in the standard way,\cite{reger} the factor $3$ accounts for 
rotational invariance in spin space. The order parameter $m_c$ defined
on a cluster will hereafter be referred to as the cluster magnetization.

The spin stiffness will also be discussed. For the non-random 2D
Heisenberg model with periodic boundary conditions it is defined as
\cite{einarsson}
\begin{equation}
\rho_s = {3\over 2}{1\over L^2}
{\partial^2 E_0(\phi)\over \partial \phi^2},
\label{rhodef1}
\end{equation}
where $\phi$ is a twist under which the interaction on all
bonds in one lattice direction is modified according to
\begin{equation}
{\bf S}_i\cdot {\bf S}_j ~\to ~ {\bf S}_i\cdot R(\phi){\bf S}_j,
\label{twist}
\end{equation}
where $R$ is the matrix rotating the $3$-component spin vector ${\bf S}_j$
by an angle $\phi$ around the spin-$z$ axis. The stiffness can be
expressed in terms of the winding number of the SSE configurations. The
winding number $W_a$, $a=x,y$, is the net number of times spin currents 
wrap around the system in the lattice direction $a$, i.e., 
\begin{equation}
W_a = (N^R_a - N^L_a)/L,
\end{equation}
where $N^R_a$ and $N^L_a$ are the number of events in the propagation with
the SSE operator sequence $S_M$ in which spin is transported  to the
``left'' and ``right'' along the $a$ direction. The winding numbers hence
take integer values $0,\pm 1,\pm 2 \ldots$. The stiffness is given by
\begin{equation}
\rho_s = {3\over 2}\langle W^2_a\rangle /\beta,
\label{rhosw2}
\end{equation}
which can be averaged over the two directions $a=x,y$.

For random systems the situation is complicated by the fact that the stiffness
can vary locally, whereas the winding number estimator (\ref{rhosw2}) is a 
global quantity characterizing the rigidity of the system as a whole (i.e.,
the energy increase due to changed boundary conditions). This global stiffness
is still an important quantity, however. One can easily prove that it is 
equivalent to an average stiffness: In a clean system, the definition 
(\ref{rhodef1}) can clearly be replaced by a definition where the twist 
(\ref{twist}) is only applied on a single boundary column (which has 
$L$ interacting pairs);
\begin{equation}
\rho_s = {3\over 2}{1\over L^2}
{\partial^2 E_0(\Phi)\over \partial \Phi^2}.
\label{rhodef2}
\end{equation}
The boundary twist here is related to the twist in the first definition
(\ref{rhodef1}) by $\Phi = L\phi$. If this definition is used for
a diluted system one still obtains the same expression (\ref{rhosw2}) in 
terms of the squared winding number, regardless of which column is taken as 
the boundary to which $\Phi$ is applied. This is because the spin currents 
wrapping around the system have to go through all $L$ columns. The number 
of interacting pairs on the boundary column can depend on which 
of the $L$ possible columns is used, however, and the currents are therefore 
distributed unequally among the bonds although the same net current passes 
through all columns. This reflects the local variation in the rigidity of 
individual bonds. The stiffness defined according to the equivalent 
definitions (\ref{rhodef1}), (\ref{rhodef2}), and (\ref{rhosw2}) is hence the 
average over all bonds of an arbitrary column. In the case that there is no 
cluster wrapping around the system in either the $x$ or $y$ direction, 
the corresponding winding number is always zero and the stiffness in that 
direction vanishes. Recent discussions of the stiffness of disordered quantum 
systems can also be found in Refs.~\onlinecite{trivedi} and 
\onlinecite{sushkov}.

The bond energy, including the constant added in (\ref{h1}), is obtained in
SSE simulations according to
\begin{equation}
E_b = -\langle H_{1,b} + H_{2,b}\rangle = -\langle n_b\rangle /\beta,
\end{equation}
where $n_b$ is the number of elements $[1,b]$ and $[2,b]$ in $S_M$. 
Hence, the average expansion power $\langle n\rangle = |E|\beta$, where $E$ is
the total internal energy. One can also show that the heat capacity 
$C=\langle n^2\rangle - \langle n\rangle^2 - \langle n\rangle$, and hence 
the width of the distribution of $n$ is $\sim \langle n\rangle ^{1/2}$
at low temperatures. This is the reason why the Taylor expansion can be 
truncated at $M \sim N\beta$.

\subsection{Convergence Issues}

In QMC studies of random systems, disorder-averaged expectation values
of the form
\begin{equation}
\langle\langle A \rangle\rangle = {1\over N_R}\sum\limits_{i=1}^{N_R} 
\langle A \rangle_i
\label{disav}
\end{equation}
normally have to be estimated using only a small subset of all $N_R$
disorder realizations. In addition, the individual expectation values
$\langle A \rangle_i$ are not evaluated exactly but are associated with 
statistical errors. Typically $\langle A \rangle_i$ is a simple operator
expectation value [such as the staggered structure factor (\ref{spi}) or
(\ref{spic})] which has an estimator that is linearly averaged over the 
importance sampled QMC configurations. In principle, the most efficient way 
to estimate the disorder average (\ref{disav}) would then be to generate only 
a single QMC configuration for each randomly selected disorder realization,
so that each term contains both sources of fluctuations (sample-to-sample 
and QMC statistical). The final
statistical error can be estimated in the standard way using data binning 
(in order to approach a Gaussian distribution from which the standard 
deviation of the average can be calculated). However, in practice this
approach is not feasible since the simulations have to be properly 
equilibrated for each disorder realization  before the QMC configurations 
can be used for averages. If a large number of MC steps are needed for 
equilibration it would clearly not be optimal to make use of only a single 
configuration. An accurate estimation of the optimum 
number of configurations would require detailed knowledge of equilibration 
times, autocorrelation times, and the statistical distributions of the 
estimators. In practice, it is rarely worthwhile to investigate 
these in detail (it would require an effort rivaling that of the actual 
simulations). In any case, the simulations should be relatively short so 
that many disorder realizations can be studied. Furthermore, the simulations
should not be dominated by equilibration. The number of MC steps used
for sampling expectation values should therefore be at least of the same 
order as the number of steps used for equilibration.

Another important issue is temperature. In order to study ground state 
properties with the SSE method, a sufficiently high inverse temperature 
$\beta$ must be used. In diluted systems, especially close to the percolation 
point, different parts of a large cluster may be connected only weakly, 
through essentially one-dimensional narrow paths 
(several examples of which can be seen in Figs.~\ref{fig:lconf} 
and \ref{fig:nconf}). Such ``weak links'' can lead to correlations that 
develop only at very low temperatures. One can therefore expect that in
order to reach the ground state much higher $\beta$ values have to be used 
than for undepleted 2D systems.

\begin{figure}
\centering
\epsfxsize=8.2cm
\leavevmode
\epsffile{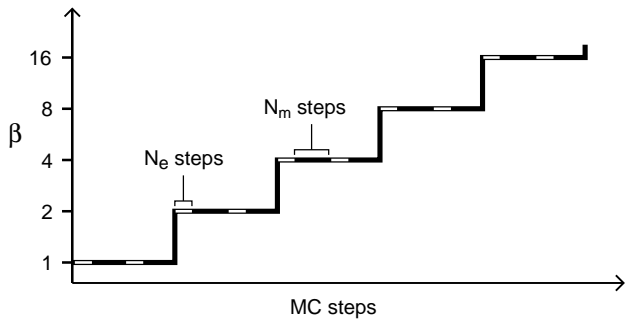}
\vskip1mm
\caption{Illustration of the $\beta$-doubling scheme used for equilibration 
and convergence to the ground state. The horizontal line segments represent 
MC steps carried out at the corresponding inverse temperatures $\beta=2^n$. 
No data is collected during the $N_e$ steps corresponding to the unfilled 
segments. Averages over the two solid segments of length $N_m$ are stored 
separately for each $\beta$.}
\label{fig:equiv}
\end{figure}

Remaining temperature effects and insufficient equilibration are two potential
sources of systematic errors in the simulations, and these have to be 
controlled very carefully. The following scheme has been developed in order 
to check for both equilibration and temperature effects: For each disorder 
realization, simulations are carried out at inverse temperatures $\beta_n 
= 2^n$, $n=0,1,\ldots, n_{\rm max}$. Starting with $n=0$ ($\beta=1$), a 
number $N_{\rm e}$ of MC steps are first carried out for equilibration.
Expectation values are sampled during the following $N_{\rm m}=2N_{\rm e}$ 
steps. At the same temperature, $N_{\rm e}$ additional steps are carried 
out during which no expectation values are sampled, again followed by 
$N_{\rm m}$ sampling steps. The second segment of $N_e+N_m$ steps is a direct 
continuation of the first one, so that the effective number of equilibration 
steps for the second sampling segment is four times that for the first one. 
A disagreement between the results of the two sampling segments then implies
that the simulation at the level of the first segment is not sufficiently 
equilibrated, and the second segment may also be affected. If the results
agree, one can conclude that at least the second segment should have 
equilibration errors that are smaller than the statistical errors (although
this should also be verified by comparing simulations with different
$N_m$, which will be done below). Since the fluctuations of the results of 
short simulations are large, the agreement between the two segments can of 
course be checked only in averages over large numbers of simulations 
of different disorder realizations.

The $\beta$-doubling scheme is illustrated in Fig.~\ref{fig:equiv}. Note that
simulations at subsequently lower temperatures can be started using the last 
SSE configuration generated at the previous temperature. An equilibrated 
configuration at $\beta$ will have an SSE sequence length $M$ approximately 
twice that in the previous run at $\beta/2$. Therefore, in order to further 
accelerate the equilibration at low temperatures, the starting sequence used 
is the previous $S_M$ doubled, i.e.,
\begin{equation}
S_{2M}=[a_1,b_1],\ldots,[a_M,b_M][a_M,b_M],\ldots,[a_1,b_1].
\label{s2m}
\end{equation}
Especially at low temperatures, where the system is almost converged to the 
ground state, the doubled SSE configuration should be very nearly distributed 
according to the equilibrium distribution at the new $\beta$. With the 
reversed order of the second set of $M$ operators in (\ref{s2m}), the initial 
$S_{2M}$ always has zero winding number, which can be expected \cite{wpaper} 
to be a slightly better starting point than the alternative one with twice 
the winding number (in practice, the difference in performance is minor).

Expectation values calculated for all $n_{max}+1$ values of $\beta$ are 
stored on disk, so that the convergence to the ground state can be checked. 
Ideally, the number of $\beta$-doublings should be large enough that there 
are no statistically significant differences between the results for 
$\beta_{max}=2^{n_{\rm max}}$ and $\beta=2^{n_{\rm max}-1}$. Since the 
asymptotic convergence is exponential, the results at  $\beta_{max}$
should then have no detectable temperature effects at the level of the 
statistical errors.

\section{Convergence tests}

In this section test results for equilibration and ground state convergence
according to the $\beta$-doubling procedures described in the previous
section are presented. Dilution fractions close to the 
percolation point can be expected to be the worst with respect to slow 
$\beta$ convergence. This is because for $p < p^*$ the largest clusters 
are two-dimensional 
and more compact than at $p^*$ (i.e., they have less ``weak links''), and 
for $p > p^*$ the cluster size does not diverge with $L$. Site diluted 
systems exactly at the percolation point are considered here.

\subsection{Equilibration}

\begin{figure}
\centering
\epsfxsize=8.2cm
\leavevmode
\epsffile{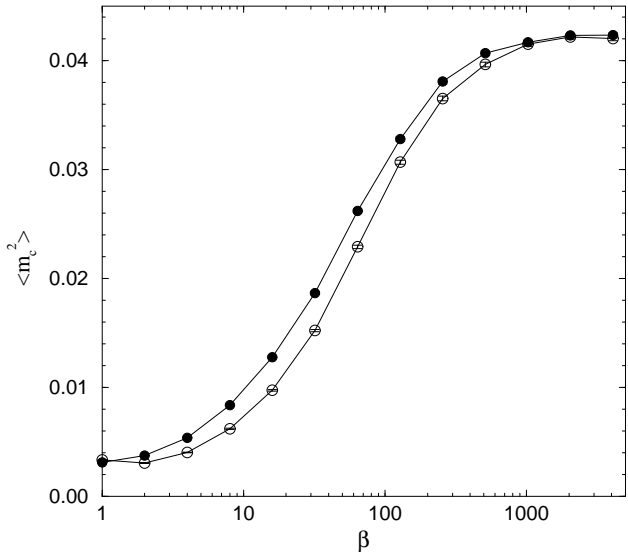}
\vskip1mm
\caption{Test results for the convergence of the sublattice magnetization of
the largest cluster on $32\times 32$ lattices at $p^*$. The number of MC steps
for averaging data for each point was $N_m=2$. The open and solid circles 
correspond to the first and second data collection segment, respectively. 
The results are averages over $10^4$ disorder realizations.}
\label{fig:m1}
\end{figure}

The equilibration of the simulations will first be 
illustrated by results for $L=32$ systems obtained with different 
$N_e$ and $N_m=2N_e$. Fig.~\ref{fig:m1} shows results for the disorder 
averaged cluster magnetization when the segments are very short; $N_e=1$ and 
$N_m=2$. At the highest temperature, $\beta=1$, the two segments give results 
that agree within statistical errors, but as the temperature is lowered the
results begin to differ considerably. At still lower temperatures the results 
again converge and become statistically indistinguishable at $\beta=1024$ in 
this case. The good agreement here can be explained by the fact that 
low-temperature simulations in the $\beta$-doubling procedure start from 
configurations which already have a rather long history at higher 
temperatures, which in combination with the trick of doubling the SSE 
operator sequence produces almost equilibrated initial configurations when 
the system is nearly in its ground state. 

\begin{figure}
\centering
\epsfxsize=8.2cm
\leavevmode
\epsffile{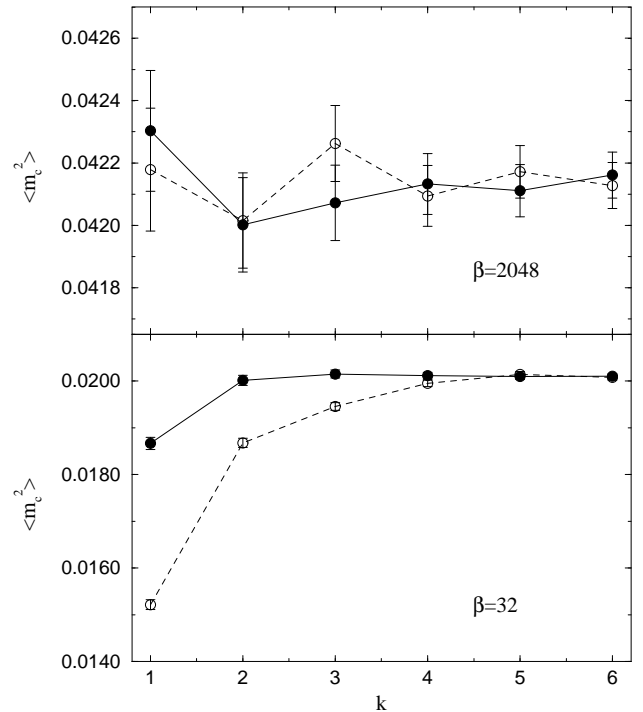}
\vskip1mm
\caption{Dependence of the calculated cluster magnetization on the number of 
MC steps in the data collection segments, $N_m=2^k$, at two different inverse 
temperatures (results for $L=32$ lattices at $p^*$, averaged over $10^4$ 
samples). The open and solid circles correspond to the first and second data 
collection segment, respectively.}
\label{fig:mk}
\end{figure}

Fig.~\ref{fig:mk} shows how results at an intermediate and low temperature
depend on the number of MC steps in the data collection segments. At
$\beta=32$, the first data segment converges after $N_m \agt 16$, whereas the 
second segment appears to be converged already for $N_m=4$. At $\beta=2048$, 
the results for the two segments agree statistically for all $N_m$, and the
averages show no discernible dependence on $N_m$. Hence, an agreement between 
the two segments indeed appears to be a good indication of sufficient 
equilibration. Since the convergence is the slowest at intermediate
temperatures, a very safe conservative check of low-temperature equilibration 
should be that the two segments agree at all temperatures. For the final 
result, the  segments can then be averaged in order to improve the statistics.
However, this typically leads only to a modest reduction of the error bars 
(i.e., significantly less than the reduction by $\sqrt{2}$ expected for 
independent data) since the statistical errors are dominated by fluctuations
between the disorder realizations. The fact that sample-to-sample fluctuations
dominate can also be seen clearly in Fig.~\ref{fig:mk}, where the error bars 
decrease much slower than by $\sqrt{2}$ for successively higher $k$.

\begin{figure}
\centering
\epsfxsize=8.2cm
\leavevmode
\epsffile{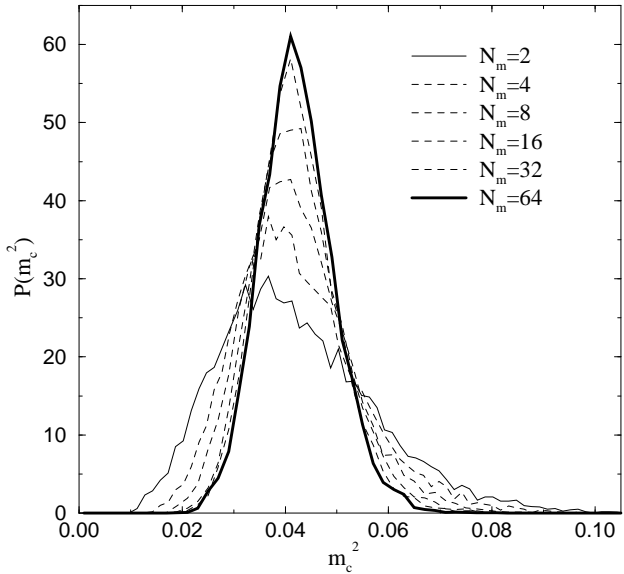}
\vskip1mm
\caption{Distribution of the cluster magnetization of $32\times 32$ lattices 
at $p^*$ for different lengths of the data collection segments. The
inverse temperature $\beta=4096$, and $10^4$ disorder realizations were 
used for each $N_m$.}
\label{fig:hist32}
\end{figure}

The results presented here indicate that even extremely short
simulations give results void of non-equilibration effects at low temperatures.
However, longer runs were used to produce some of the data presented
in this paper. The main reason for this is that although unbiased disorder 
averages of the form (\ref{disav}) can be obtained with short simulations, 
large statistical errors in the individual expectation values can be 
problematic when considering non-linear functions of the expectation values 
(such as their typical values) or their complete statistical distributions. 
One then has to demand that the statistical errors of the individual 
expectation values are much smaller than the width of the distribution of
the exact expectation values. An example of how statistical errors can distort
distributions is shown in Fig.~\ref{fig:hist32}, where histograms of the 
cluster magnetization are compared for six different simulation lengths. 
Both the data collection segments were used for calculating the individual 
expectation values, i.e., the number of measurements for each realization is 
$2N_m$. The histograms become significantly narrower as the number of MC 
steps is increased. The distribution is not completely converged even for 
the longest simulation considered here ($N_m=64$), but the relatively small 
differences between $N_m=32$ and $64$ suggest that the $N_m=64$ result is 
close to the exact distribution. Note that the first moment of the 
distribution, i.e., the linear disorder average (\ref{disav}), is the same 
within statistical errors for all $N_m$ (which was demonstrated at 
$\beta=2048$ in Fig.~\ref{fig:mk}).

In the calculations discussed in the following sections, $N_m$ between $100$ 
and $250$ was used to ensure that reliable distributions could be obtained
at the percolation point. For $p < p^*$, where the full distributions are
not as important, $N_m=50$ was typically used. Since effects of insufficient 
equilibration are undetectable even in much shorter simulations the results 
should definitely be void of any bias of this nature. 

\subsection{Ground state convergence}

\begin{figure}
\centering
\epsfxsize=8.4cm
\leavevmode
\epsffile{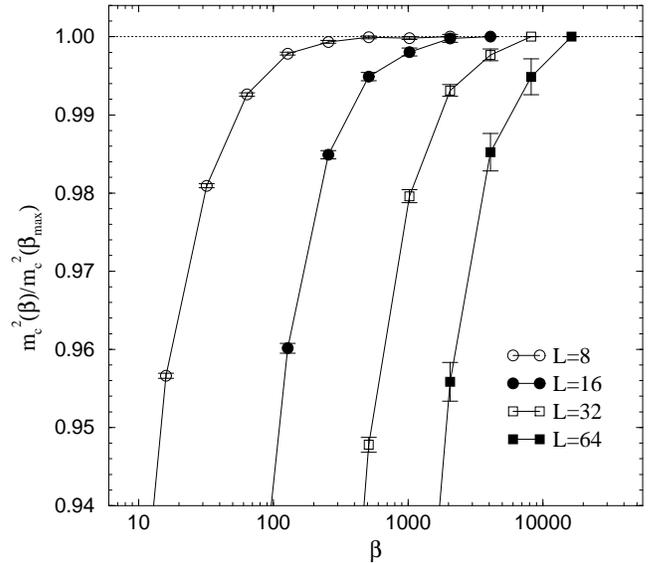}
\vskip1mm
\caption{Cluster magnetization ratios vs inverse temperature for $L\times L$ 
lattices. The number of samples used for averaging the results was 
$88000$, $21000$, $9000$, and $2500$, for $L=8,16,32$, and $64$, 
respectively.}
\label{fig:ratl}
\end{figure}

Already the results shown in Fig.~\ref{fig:m1} demonstrate that very low 
temperatures are required in order to converge the sublattice magnetization 
to its ground state value. For $L=32$, a $\beta$-value higher than $2000$ is 
needed to eliminate temperature effects within the statistical errors.
In order to more accurately study remaining effects at 
low temperatures it is useful to calculate ratios 
$\langle m_c^2(\beta_i)\rangle / \langle m_c^2(\beta_j)\rangle$ of the squared
cluster magnetization at different temperatures. The relative statistical 
errors are smaller in the ratios than in the absolute values, since the 
sample-to-sample fluctuations cancel when the same realizations are used at 
all temperatures. Fig.~\ref{fig:ratl} shows results for systems with 
$L=8,16,32$ and $64$, which were simulated with $\beta$ up to $\beta_{\rm max}
= 256 \times L$. The ratios, with the data at the respective $\beta_{\rm max}$
in the denominator, were analyzed using the bootstrap method \cite{bootstrap}
in order to obtain accurate estimates of the error bars. For $L=8$ and $L=16$,
the results at $\beta_{\rm max}$ and $\beta_{\rm max}/2$ do not differ within 
statistical errors and hence the result at $\beta_{\rm max}$ should not have 
any temperature effects left at this precision level. The $L=32$ and $64$ 
results are not completely converged to the ground state, however. The 
exponential low-temperature convergence seen for all the system sizes 
indicates that the remaining temperature effects at $\beta_{\rm max}$ should 
only lead to an error that is smaller than the difference between the ratios 
at $\beta_{\rm max}$ and $\beta_{\rm max}/2$. Hence, the under-estimation of 
the sublattice magnetization should be less than $0.2$\% for $L=32$ and less 
than $0.5$\% for $L=64$. These upper bounds for the systematic errors are 
of the same magnitude as the respective statistical errors in $\langle 
m^2_c\rangle$ (which unlike the ratios also include contributions from 
sample-to-sample fluctuations). The remaining small temperature effects should
therefore not substantially affect the finite-size scaling of the sublattice 
magnetization (to be discussed in the next section).

Fig.~\ref{fig:ratl} shows magnetization ratios for fixed-$N_c$ clusters,
with $\beta_{\rm max}=32\times N_c$. In this case there are small but
detectable differences between the results at $\beta_{\rm max}$  and
$\beta_{\rm max}/2$ for all system sizes, except for $N_c=1024$ where 
the statistical error is larger than the difference. Again, the maximum
relative systematic errors remaining at $\beta_{max}$ are similar in 
magnitude to the statistical fluctuations in $\langle m_c^2\rangle$ and can 
only have very minor effects on the finite-size scaling. 

The $\beta$ needed
for ground state convergence decreases rapidly away from the percolation
point, and therefore the $p<p^*$ results for $L\times L$ lattices discussed 
in Secs.~V and VI are completely converged even for $L=64$.

\section{Long-range order in percolating clusters}

In this section the ground state sublattice magnetization of the percolating 
cluster is investigated in detail. If it remains finite in the thermodynamic
limit, the order-disorder transition driven by dilution must necessarily occur
exactly at the classical percolation density. To see this, consider the 
sublattice magnetization (\ref{subm}) of the diluted $L\times L$ lattice. 
Its disorder average can be written as a sum of contributions from all the 
clusters $k$ on the lattices as,
\begin{equation}
\langle m^2 \rangle= 
{1\over N^2} \left\langle \sum_k N_k^2 m_k^2 \right\rangle.
\end{equation}
In the thermodynamic limit, only infinite clusters contribute to this sum,
and therefore one only needs to consider the behavior of the cluster
magnetizations $m_k^2$ for large clusters. If there is long-range order, it 
is natural to assume that the sublattice magnetization is 
self-averaging (a fact that will also be demonstrated explicitly below). 
The individual $m_k^2$ values can then be replaced by the infinite-size
extrapolated average for the largest cluster, i.e., $\langle m_c^2\rangle$,
which gives
\begin{equation}
\langle m^2 \rangle =  
{ \langle m_c^2\rangle \over N^2} \left\langle \sum_k N_k^2 
\right\rangle , ~~~~~ (L\to \infty).
\label{infmag}
\end{equation}
This expression is identical to the order parameter of a classical diluted 
system, up to the factor $\langle m_c^2\rangle$ which is reduced by quantum
fluctuations from its classical value $1/4$ (for an Ising model with
$S_i^z=\pm 1/2$). If $\langle m_c^2\rangle$ remains finite at $p = p^*$
[which is the condition for (\ref{infmag}) to remain valid for all
$p \le p^*$] the only singular behavior is in the classical expectation 
value and hence the critical behavior is that of classical percolation.

\begin{figure}
\centering
\epsfxsize=8.4cm
\leavevmode
\epsffile{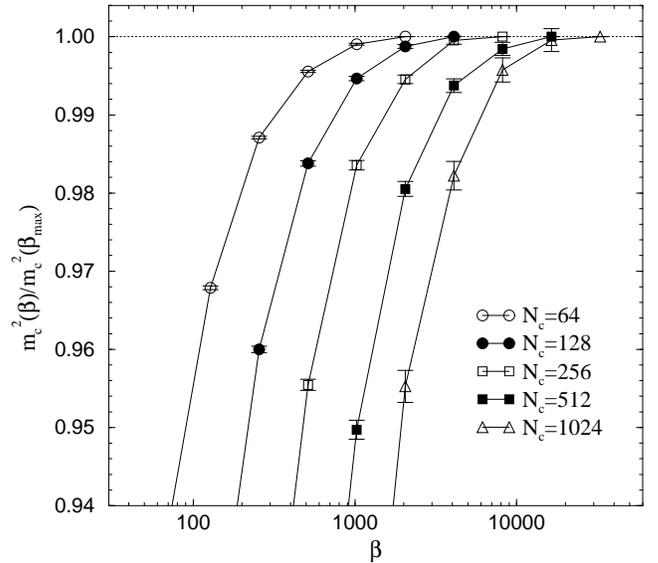}
\vskip1mm
\caption{Magnetization ratios vs inverse temperature for fixed-$N_c$ 
clusters. The number of samples was $43000$, $15000$, $10000$, $3000$, 
and $1100$, for $N_c=64,128,256,512$, and $1024$, respectively.}
\label{fig:ratn}
\end{figure}

In general, Eq.~(\ref{infmag}) holds for any dilution fraction $p < p_c$,
where $p_c$ in principle may be less than $p^*$. The cluster magnetization 
$\langle m_c^2\rangle ^{1/2}$ will here be determined at percolation, where 
the infinite clusters are fractal. Dilutions less than the percolation 
density, where the infinite clusters are two-dimensional, will be studied 
in the next section.

\subsection{Sublattice magnetization in site-diluted systems}

At the percolation point, the average number of spins in the largest cluster 
on a diluted $L\times L$ lattice scales asymptotically as $\langle N_c 
\rangle \sim L^d$, with $d$ the fractal dimension $91/48$.\cite{stauffer} 
As can be seen in Fig.~\ref{fig:nc}, the full distribution of the size of 
the largest cluster  also scales as $L^d$, i.e., the distribution width 
also diverges as $L \to \infty$. This is in sharp contrast to the situation 
below the percolation threshold where the size distribution approaches a 
$\delta$-function at a size $\sim L^2$.  Note, however, that the scaled 
distribution at $p^*$ has sharp cut-offs both at the lower and upper edge, 
meaning that also the smallest and largest clusters grow as $L^d$. 
Hence, finite-size scaling of 
$\langle m_c^2\rangle$ calculated on such fluctuating-$N_c$ clusters as a 
function of $L$ is a well defined procedure for extracting the sublattice 
magnetization of the infinite fractal cluster. Nevertheless, the alternative 
way of approaching the thermodynamic limit with fixed-$N_c$ clusters on
the infinite lattice is also considered here. An agreement between the two 
calculations will provide additional support to the argument \cite{awscomment}
that the percolating cluster is ordered.

\begin{figure}
\centering
\epsfxsize=8.2cm
\leavevmode
\epsffile{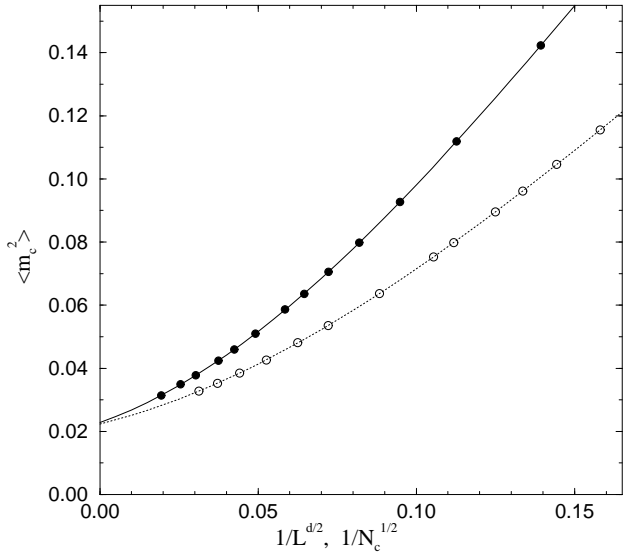}
\vskip1mm
\caption{Finite-size scaling of the disorder averaged squared cluster 
magnetization. The results for the largest cluster on $L\times L$ lattices
(solid circles) are graphed vs $L^{-d/2}$ and those for fixed-$N_c$
clusters (open circles) vs ${N_c}^{-1/2}$. Statistical errors are much
smaller than the symbols. The curves are cubic polynomial fits.}
\label{fig:magln}
\end{figure}

In the pure 2D Heisenberg model the leading size-correction \cite{huse}
to $m^2$ is $\sim N^{-1/2}$, which can be seen clearly in numerical data. 
\cite{reger,sse2} In analogy with this, as a scaling hypothesis at 
percolation, the following leading size corrections are tested here 
for the largest cluster on $L\times L$ lattices and fixed-$N_c$ clusters,
respectively:
\begin{eqnarray}
\langle m_c^2\rangle_L & = & \langle m_c^2\rangle_\infty + a L^{-d/2}, 
\label {scalel} \\
\langle m_c^2\rangle_{N_c} & = & \langle m_c^2\rangle_\infty + b N^{-1/2}_c.
\label{scalen}
\end{eqnarray}
Fig.~\ref{fig:magln} shows results for $L$ up to $64$ and $N_c$ up to $1024$. 
The data are fully consistent with the scaling ansatz, although 
in order to fit all the points a polynomial cubic in $L^{-d/2}$ has to be 
used in both cases (a cubic polynomial is needed also to fit high-accuracy 
data for the clean 2D Heisenberg model\cite{sse2}). The infinite-size 
extrapolated values for $\langle m_c^2 \rangle$ from the two fits agree 
very well (within statistical errors). The sublattice magnetization is 
in fact quite large, $\langle m_c \rangle = 0.150(2)$, which is almost 
precisely half of the value $m = 0.307$ for the clean 2D system.
\cite{reger,sse2} 

It should be stressed that it is not critical whether or not the scaling 
ansatz assumed here to carry out the extrapolation of the sublattice
magnetization is strictly correct or not. Unless the behavior would change 
dramatically for even larger systems, a slightly different finite-size 
correction would not significantly affect the extrapolated $\langle m_c^2 
\rangle$. One could of course argue that a cross-over to a qualitatively
different behavior cannot be excluded, as indeed has been done.\cite{todo} 
No plausible physical reason for such a cross-over has been presented, 
however. With two different boundary conditions for the clusters giving the 
same result for the infinite-size extrapolated sublattice magnetization, 
the most natural scenario must be that the percolating cluster is ordered. 

\begin{figure}
\centering
\epsfxsize=8.2cm
\leavevmode
\epsffile{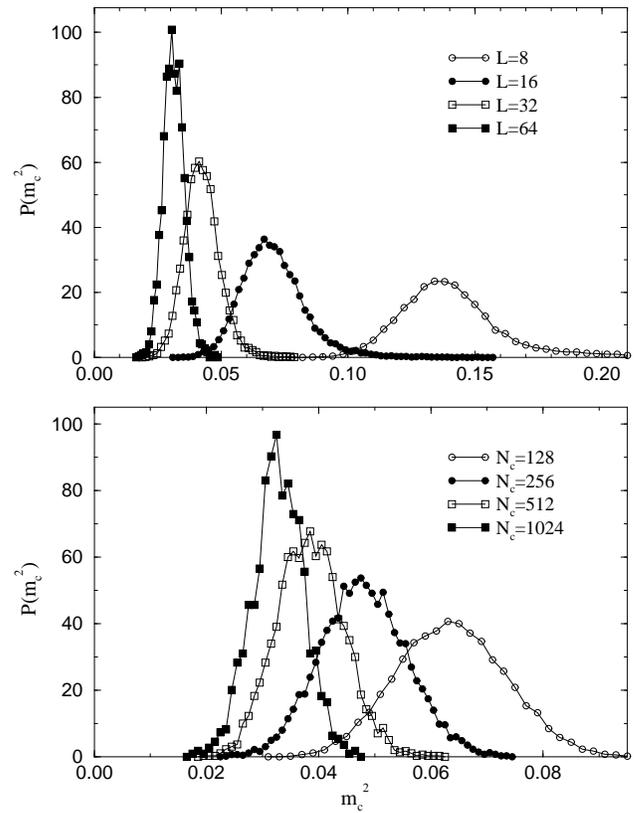}
\vskip1mm
\caption{Distribution of the cluster magnetization of $L\times L$
lattices (top panel) and fixed-$N_c$ clusters on the infinite lattice 
(bottom panel).}
\label{fig:histln}
\end{figure}

\begin{figure}
\centering
\epsfxsize=8.2cm
\leavevmode
\epsffile{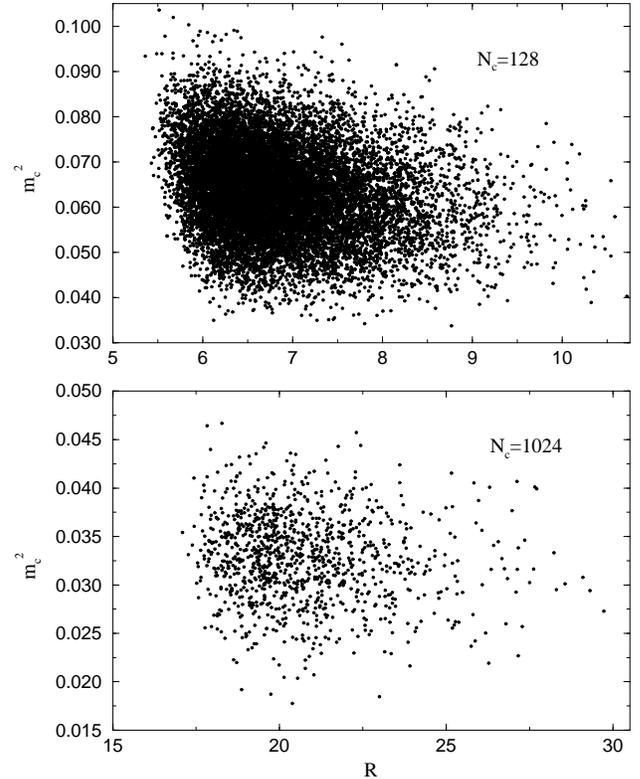}
\vskip1mm
\caption{Individual squared magnetizations vs the radius of gyration
for clusters on the infinite lattice.}
\label{fig:gyr}
\end{figure}

In a disordered system the order parameter is not constant over the whole
system, but depends locally on the structure of the lattice. One would, 
however, expect self-averaging, i.e., the sublattice magnetization averaged 
over different regions of an infinite cluster should be the same when the size 
of the regions is sufficiently large. In finite systems, self-averaging 
can be seen in the statistical distributions of the individual cluster 
magnetizations. Fig.~\ref{fig:histln} shows results at the percolation
point for several $L\times L$ and fixed-$N_c$ systems. As discussed in 
Sec.~III, the histograms can be expected to be slightly broadened by the 
statistical fluctuations in the SSE results for the individual $m^2_c$ values.
Such effects should, however, be minor when the simulations are as long as 
those used for the data shown here ($N_m=100$ for the $L\times L$ lattices and
$250$ for the fixed-$N_c$ clusters). The widths of both types of distributions 
decrease with increasing system size, which is consistent with vanishing
fluctuations in the thermodynamic limit. It can also be noted that the
distributions become more symmetric for larger systems --- the weak
tails visible at the high-end of the distributions for small clusters 
vanish as the system grows. The behavior is hence fully consistent with the 
$\delta$-function distribution expected for a self-averaging quantity in 
the thermodynamic limit.

It is also interesting to study how the cluster magnetization depends on
the shape of the cluster. A compact cluster is likely to have a stronger
order than one which has many narrow paths. A natural length scale 
characterizing the over-all density of the unconstrained fixed-$N_c$ 
clusters is the radius of gyration, 
\begin{equation}
R = \left ( {1\over 2N_c^2}
\sum\limits_{i=1}^{N_c} \sum\limits_{j=1}^{N_c} 
(x_i - x_j)^2 + (y_i - y_j)^2 \right )^{1/2},
\end{equation}
where $x_i,y_i$ are the (integer) coordinates of the magnetic sites. 
Fig.~\ref{fig:gyr} shows scatter plots of the cluster magnetization
versus $R$ for two cluster sizes. For $N_c=128$, one can see that the most 
compact clusters, i.e., those with the smallest $R$, indeed have the largest 
magnetizations. After an initial rapid decrease with $R$ for the smallest
$R$, the average magnetization only decreases slowly with increasing $R$, 
however. The $N_c=1024$ clusters show a similar behavior. There are of course
in principle clusters with very large $R$ that should have much smaller 
magnetizations, but these clusters lack statistical significance. The weak 
$R$-dependence for the statistically significant clusters is another 
manifestation of a strongly self-averaging sublattice magnetization.

\subsection{Sublattice magnetization in bond-diluted systems}

\begin{figure}
\centering
\epsfxsize=8.2cm
\leavevmode
\epsffile{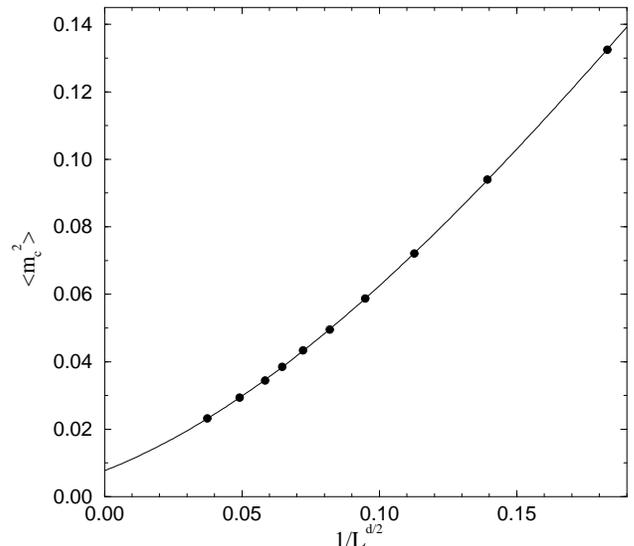}
\vskip1mm
\caption{Finite-size scaling of the disorder averaged squared cluster 
magnetization of the bond diluted system at the percolation density. 
The curve is a cubic polynomial fit.}
\label{fig:magb}
\end{figure}

For the bond-diluted system only simulations of $L\times L$ lattices were
carried out. Fig.~\ref{fig:magb} shows the results for the cluster 
magnetization at the bond percolation point, $P^*=1/2$, graphed in the same 
way as for the site diluted systems in Fig.~\ref{fig:magln}. Also in this case
the scaling to a finite sublattice magnetization is evident, but the value of 
the order parameter is smaller than in the site diluted system; 
$\langle m_c\rangle = 0.088(2)$. The difference can be explained by the 
different local structures of the two types of lattices. Although the fractal 
dimension $d$ of the cluster is the same for site and bond percolation,
\cite{marro} the average number of bonds per spin is smaller in the bond 
diluted case --- $1.121$  versus $1.259$. This leads to stronger quantum 
fluctuations in the bond-diluted system. The infinite-size energy per bond 
(which reflects the tendency to nearest-neighbor singlet formation) is 
$-0.3890(1)$ and $-0.4068(2)$ for site and bond dilution, respectively.

It can also be noted that for a given $L$ the average largest cluster on
the bond diluted lattice is $\approx 45$\% larger than on the site diluted
lattice. The stronger quantum fluctuations and the larger cluster sizes 
imply that for given $L$ a lower temperature has to be used to converge 
the bond diluted system to the ground state. For the largest size studied in 
this case, $L=32$, an inverse temperature $\beta=32768$ was used.

\subsection{Scaling of the full staggered structure factor}

\begin{figure}
\centering
\epsfxsize=8.2cm
\leavevmode
\epsffile{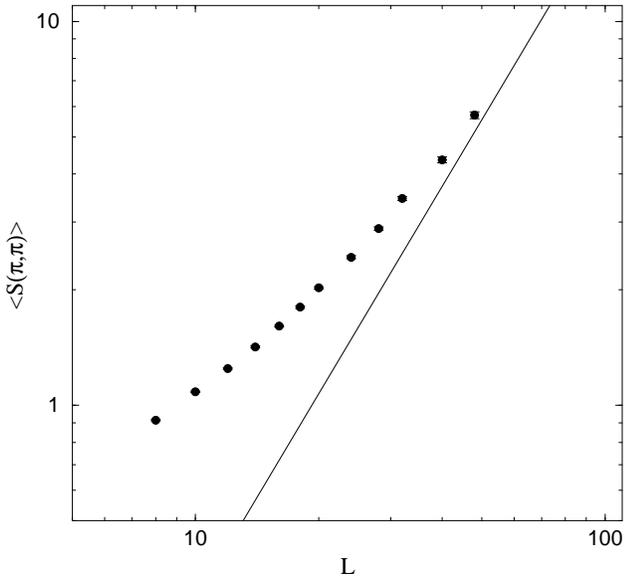}
\vskip1mm
\caption{Finite-size scaling of the disorder averaged staggered structure 
factor of the full site diluted $L\times L$ lattice. The line has slope 
$43/24$, expected for classical percolation.}
\label{fig:spi}
\end{figure}

The previous claims of quantum criticality at the percolation point
\cite{kato,yasuda2} were primarily based on a finite-size scaling analysis of 
the staggered structure factor. A log-log plot of $S(\pi,\pi)$ calculated 
using SSE simulations including all the spins of diluted $L\times L$ 
lattices is shown in Fig.~\ref{fig:spi}. The numerical values agree well 
with those of Ref.~\onlinecite{kato}. One can, however, expect a barely
discernible finite-$T$ reduction in the previous $L=48$ results because the 
temperature used ($\beta=1000$) was not sufficiently low for complete 
converge to the ground state (see Fig.~\ref{fig:ratl} and a related 
discussion in Ref.~\onlinecite{awscomment}). 

The scaling seen in Fig.~\ref{fig:spi} is different from the expected 
classical percolation behavior. Given the results presented above for the 
scaling of the cluster magnetization, the deviation from classical behavior 
for this range of system sizes is not surprising, however. Classically, the 
finite-size scaling of $S(\pi,\pi)$ is solely the result of the divergence 
of the size of the connected clusters with $L$. In the quantum mechanical 
case, there is also a factor, the sublattice magnetization of the cluster, 
multiplying each cluster size, i.e.,
\begin{equation}
\langle S(\pi,\pi) \rangle = {1\over N}
\left\langle\sum_k N_k^2 m_k^2 \right\rangle .
\label{spiqm}
\end{equation}
The size-dependence of the average $m_k^2$ was shown in Fig.~\ref{fig:magln}.
From these results it is clear that there is an effect that partially 
compensates for the growth of the cluster sizes $N_k$ in Eq.~(\ref{spiqm}), 
namely, $m_k^2$ decreases with increasing cluster size. Hence, for systems
where the relative size corrections to the cluster magnetization are 
still significant, as is the case for all sizes that can currently be 
reached in numerical simulations, the growth of $S(\pi,\pi )$ with $L$ will 
be slower than for a classical system. This explains the slow convergence 
towards the classical behavior that can be seen in  Fig.~\ref{fig:spi}. 
It can be noted that the largest cluster completely dominates the staggered 
structure factor and the curve shown in Fig.~\ref{fig:spi} changes only very 
little if only the largest cluster is included, i.e., $S_c(\pi,\pi) 
\approx S(\pi,\pi)$.\cite{todonote}

\section{Dilution dependence of the sublattice magnetization}

\begin{figure}
\centering
\epsfxsize=8.2cm
\leavevmode
\epsffile{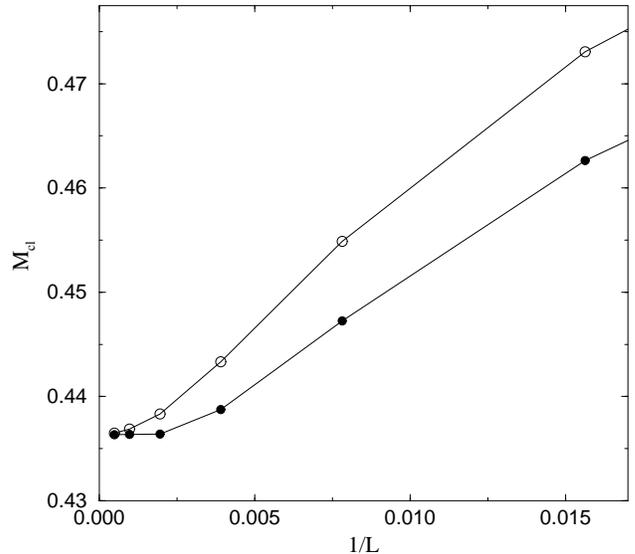}
\vskip1mm
\caption{Size dependence of the classical magnetization for systems 
diluted at one half percent less than the percolation density ($p=p^*-0.005$). 
The open circles correspond to the full cluster sum, 
Eq.~(\protect{\ref{mcl1}}). The solid circles are from the average 
including only the largest cluster, Eq.~(\protect{\ref{mcl2}}).}
\label{fig:mcl}
\end{figure}

The doping dependence of the sublattice magnetization of antiferromagnetic
cuprates can be measured experimentally using NQR, $\mu ^+$SR and
neutron scattering.\cite{doping,greven} 
Results for the Heisenberg model were recently obtained using 
an improved spin-wave theory which, however, breaks down close to the
percolation threshold (the critical point is unphysical, located at a hole 
concentration {\it higher} than the percolation density).
\cite{chernyshev,chernyshev2} Previous QMC calculations of the doping 
dependence were based on Eq.~(\ref{subm}).\cite{kato} Use of this formula 
becomes very difficult close to the percolation threshold, however, since the 
smallness of the sublattice magnetization there is associated with a slow 
convergence to the asymptotic regime in which finite-size scaling is reliable.
Here a different approach will be taken, based on the fact that the sublattice 
magnetization can be decomposed into classical and quantum mechanical 
factors, which can be evaluated separately. This decomposition was already 
discussed in Sec.~IV and was written down as Eq.~(\ref{infmag}). Here the 
notation $M=\langle m^2 \rangle ^{1/2}$ will be used for the disorder 
averaged sublattice magnetization. Eq.~(\ref{infmag}) can then be written as
\begin{equation}
M(p) = M_{qm}(p)M_{cl}(p), ~~~~~ (L\to \infty),
\label{mdecomp}
\end{equation}
where $M_{qm}$ is the quantum mechanical factor
\begin{equation}
M_{qm} = \sqrt{\langle m_c^2\rangle},
\label{mqm}
\end{equation}
and $M_{cl}$ is the classical (geometrical) factor
\begin{equation}
M_{cl} = {1\over N}\left\langle \sum_k N_k^2 \right\rangle^{1/2}.
\label{mcl1}
\end{equation}
In the ordered regime, $0 \le p < p^*$, only the largest cluster 
contributes to this sum in the thermodynamic limit. The classical factor 
can therefore also be obtained as
\begin{equation}
M_{cl} = \langle N_c\rangle /N .
\label{mcl2}
\end{equation}

Fig.~\ref{fig:mcl} shows the size convergence using the two definitions of the 
classical factor when the dilution fraction $p=p^*-0.005$. The single-cluster 
average (\ref{mcl2}) clearly converges faster. It is apparent that a reliable 
extraction of the quantum mechanical sublattice magnetization $M$ using the 
structure factor formula (\ref{subm}) would be impossible in this case, 
since not even the classical magnetization is in the asymptotic scaling 
regime for the range of system sizes where QMC simulations can be carried out 
($L \alt 100$). The quantum mechanical $M$ can be expected to have an even 
worse scaling behavior, due to effects similar to those found for the 
staggered structure factor in Sec.~IV-C. The quantum mechanical factor 
$M_{qm}$ can be calculated based on much smaller system sizes, however. 
It was evaluated in the extreme case $p=p^*$ in Sec.~IV, and even there it 
is as large as $50\%$ of the value in the other extreme, i.e., the 
non-diluted system ($p=0$). Hence, $M_{qm}$ is only weakly dependent on 
the dilution fraction, and most of the $p$ dependence of $M$, including 
the singular behavior at $p^*$, is in the classical factor $M_{cl}$.

\begin{figure}
\centering
\epsfxsize=8.2cm
\leavevmode
\epsffile{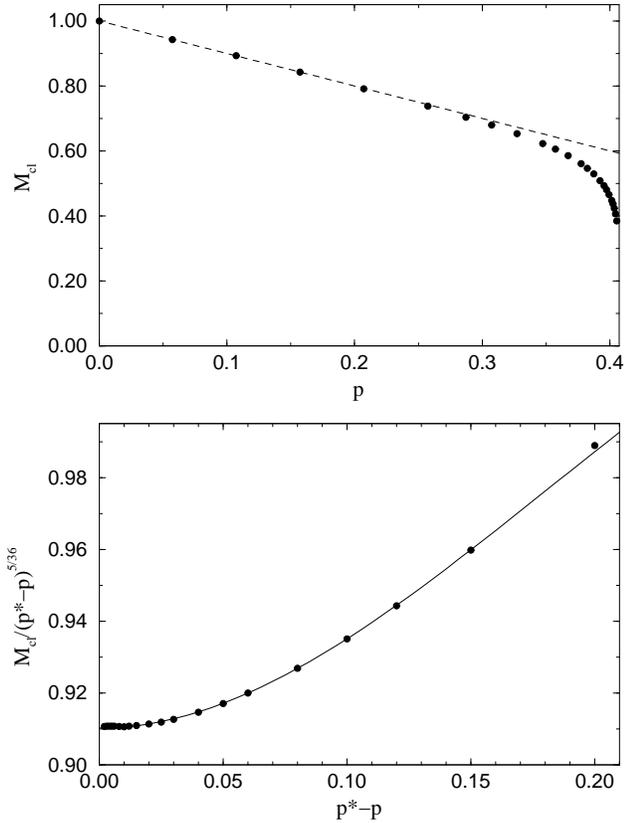}
\vskip1mm
\caption{Upper panel: Classical magnetization vs dilution. The dashed line
shows the small-$p$ form $M_{cl} = 1-p$. Lower panel: The magnetization 
divided by $(p^*-p)^{5/36}$ graphed vs $p^*-p$. The curve is a polynomial
fit, with parameters given in Eq.~(\protect{\ref{mclb}}).}
\label{fig:magcl}
\end{figure}

The classical magnetization is known to vanish at the percolation threshold 
with the exponent $5/36$,\cite{stauffer} i.e.,
\begin{equation}
M_{cl}(p \to p^*) = A_{cl}(p^*-p)^{5/36}.
\label{mclexp}
\end{equation}
In the weak dilution limit, one can easily obtain the result $M_{cl}= 1-p$.
Numerical values for $0 \le p \le p^* -0.002$ were obtained here
by simulations of lattice sizes as large as $L=4096$, using the single-cluster 
estimator (\ref{mcl2}). In the example illustrated in Fig.~\ref{fig:mcl}, the
results for the three largest sizes are $0.43640(4)$, $0.43636(3)$ , and
$0.43634(2)$ (for $L=512$, $1024$, and $2048$, respectively) and the 
$L=2048$ result (which is based on $3\times 10^5$ samples) can hence be taken 
as the infinite-size value of $M_{cl}(p^*-0.005)$. Closer to the percolation 
threshold $L=4096$ lattices were used. Fig.~\ref{fig:magcl} shows results for 
the whole dilution range. The linear small-$p$ form describes the data well 
for $p$ up to $\approx 0.2$. The asymptotic form (\ref{mclexp}) is well 
reproduced for $p^*-p \alt 0.02$, with the constant $A_{cl}\approx 0.91$. In 
order to have an analytic expression describing $M_{cl}$ in a wider region 
around $p^*$, higher-order terms can be added to $A_{cl}$. The following 
forms will be used in combination with fits to the quantum mechanical factor 
in order to obtain expressions for $M$ both close to 
$p=0$ and $p=p^*$:
\begin{eqnarray}
M_{cl}(p \alt 0.2) & = & 1 - p, \label{mcla} \\
M_{cl}(p^*-p \alt 0.2) & = & [0.9102+3.053(p^*-p)^2 \nonumber \\
                      &&  -5.642 (p^*-p)^3)](p^*-p)^{5/36}. 
                           \label{mclb}
\end{eqnarray}
Note that it is not claimed here that the higher-order terms in the form
(\ref{mclb}) are the correct subleading terms of the critical percolation 
behavior --- the purpose is just to have an expression that describes the 
data well in practice, within the stated region.

\begin{figure}
\centering
\epsfxsize=8.2cm
\leavevmode
\epsffile{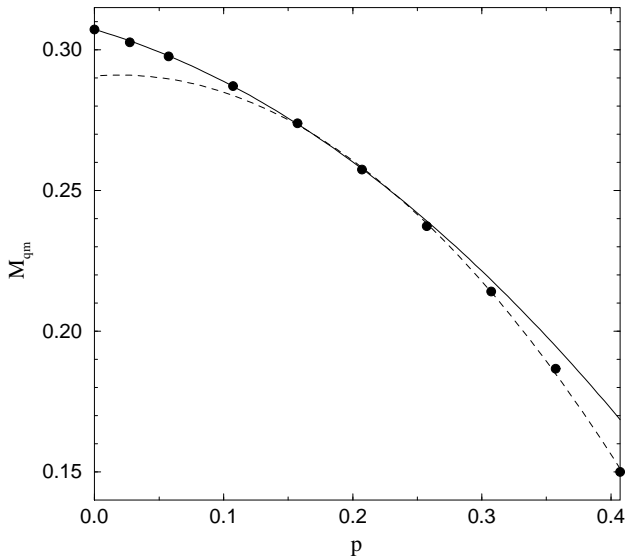}
\vskip1mm
\caption{The quantum mechanical factor (cluster magnetization) vs dilution.
The curves are quadratic fits (solid for small $p$ and dashed for
small $p^*-p$) with parameters given in Eq.~(\protect{\ref{mqma}}) and 
(\protect{\ref{mqmb}}), respectively.}
\label{fig:mqm}
\end{figure}

The quantum mechanical factor can be calculated in the same way as was already
explained in the case of $p=p^*$ in Sec.~IV, i.e., using the SSE method for
the largest cluster of connected magnetic sites on $L \times L$ lattices. 
Here $L$ up to $64$ was used for dilution fractions $p=p^* - \delta$, with
$\delta = 0.05,0.10,\ldots, 0.35$, and $0.38$. SSE simulations at $p=0$ were 
previously carried out for $L$ up to $16$,\cite{sse2} and were here extended 
up to $L=64$. Since the ground state convergence occurs at lower $\beta$ 
as $\delta$ is increased, the simulations are faster than at $p^*$ and
a larger number of samples could therefore be studied. The nuber of samples 
was $> 10^4$ for $L=64$ and up to $10^6$ for smaller lattices. The results 
were extrapolated to infinite size using a leading correction $\sim 1/L$ 
(as in the case of the clean 2D system\cite{reger}). The resulting 
$M_{qm}(p)$ is shown in Fig.~\ref{fig:mqm}, along with two quadratic 
fits which describe the data very well over quite wide ranges of $p$. 
The fitted forms are
\begin{eqnarray}
M_{qm}(p \alt 0.25) & = & 0.3072-0.134\cdot p-0.51\cdot p^2, \label{mqma} \\
M_{qm}(p^*-p \alt 0.25) & = & 0.151+0.721\cdot (p^*-p) \nonumber \\
                         && -0.93\cdot (p^*-p)^2. \label{mqmb}
\end{eqnarray}

\begin{figure}
\centering
\epsfxsize=8.2cm
\leavevmode
\epsffile{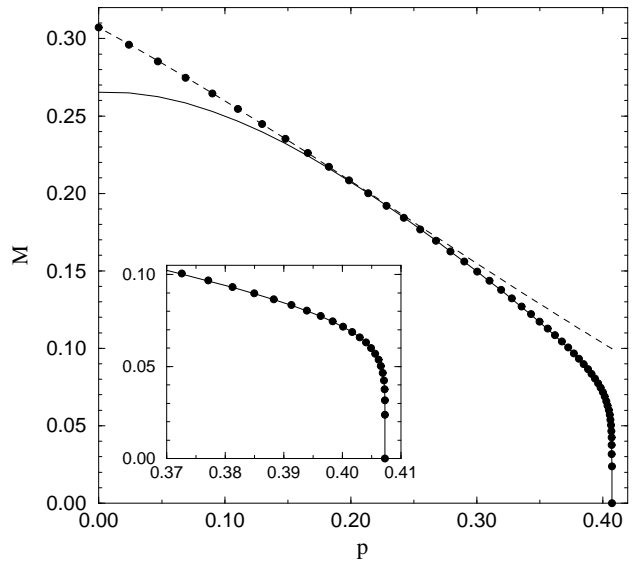}
\vskip1mm
\caption{Sublattice magnetization vs site-dilution (solid circles). The 
curves are parametrized forms discussed in the text. The inset shows the 
behavior close to the percolation threshold on a more detailed scale.}
\label{fig:mfull}
\end{figure}

The final result for the sublattice magnetization of the site diluted 
Heisenberg model is shown in Fig.~\ref{fig:mfull}. The solid circles were 
obtained by interpolating the numerical results for $M_{cl}$ and $M_{qm}$ 
and multiplying them according to Eq.~(\ref{mdecomp}). Forms describing the 
results well in quite wide regions $p \alt 0.25$ and $p^*-p \alt 0.25$ can be 
obtained by multiplying the corresponding expressions (\ref{mcla}), 
(\ref{mqma}) and (\ref{mclb}), (\ref{mqmb}). The resulting curves are also 
shown in the figure. The initial linear reduction $M(p)/M(0) = 1-B_0p$, where 
the coefficient and its estimated error is $B_0=1.44 \pm 0.05$. At the 
percolation threshold the leading behavior is $M(p) = A_*(p^*-p)^{5/36}$ with
$A_* = 0.137 \pm 0.002$.

In Ref.~\onlinecite{chernyshev}, the sublattice magnetization normalized by 
the number of magnetic sites, $M'(p)=M(p)/(1-p)$, which is equivalent to
the quantum mechanical factor $M_{\rm qm}$ for small $p$, 
was calculated using spin 
wave theory with a T-matrix approximation. The initial linear weak-dilution 
form $M'(p)/M'(0) = 1-B'_0p$, with $B'_0 = 0.691 \pm 0.005$, was found in that 
approximation. The results obtained here for $M(p)$ correspond to a 
slightly smaller coefficient; $B'_0 = 0.44 \pm 0.05$. Despite this
difference in initial slope, the relative agreement between the full
sublattice magnetization shown Fig.~\ref{fig:mfull} and the corresponding
spin-wave result is very good up to $p \approx 0.15$ (not shown here---
see Fig.~10 of Ref.~\onlinecite{chernyshev2}). For higher $p$,
the spin-wave result falls significantly below the QMC result until very close 
to the percolation point, where the actual $M(p)$ approaches zero but the
spin-wave result remains finite.\cite{chernyshev2} For $p \le 0.35$, the 
results shown in Fig.~\ref{fig:mfull} agree well with those presented 
previously by Kato {\it et al}.\cite{kato} Their 
extrapolations closer to the percolation threshold are not reliable, however, 
since they fitted a different, non-classical exponent to describe the 
$p \to p^*$ behavior. The estimated accuracy for the $M(p)$ curve obtained 
here is better than 2\% over the whole range of dilutions (significantly
better for $p \alt 0.1$).

\section{Spin stiffness}

As discussed in Sec.~II-B, the spin stiffness $\rho_s$ can be obtained in SSE 
simulations in terms of the winding number fluctuations, Eq.~(\ref{rhosw2}).
At dilution fractions $p > p^*$ there are no clusters wrapping around the 
periodic $L\times L$ lattice for large $L$, and therefore the stiffness 
vanishes identically. Exactly at the percolation threshold the wrapping 
probability in a given direction is approximately $0.52$,\cite{percdens} 
and the stiffness can then be non-zero for finite $L$. It should vanish 
in the thermodynamic limit, however. For $p < p^*$ the wrapping probability
approaches $1$ as $L \to \infty$ and in view of the existence of
antiferromagnetic long-range order the stiffness can then be expected
to be non-zero.

For the classical diluted Heisenberg model the behavior of the stiffness
(or helicity modulus) is known.\cite{harris1} It scales in the same way as
the conductivity of a random resistor network, with the conductivity
exponent $t$ of percolation.\cite{harris2} According to recent simulations, 
the value of this exponent is $t = 1.310(1)$.\cite{grassberger} With the 
long-range order present in the percolating clusters of the quantum Heisenberg
model, as demonstrated in Sec.~IV, one can expect that the stiffness should 
behave as in the classical model (in analogy with the ``renormalized 
classical'' behavior of the clean 2D Heisenberg model,\cite{chn} and in view 
of general symmetry arguments). Scaling with the conductivity exponent will 
therefore be tested here. It can be noted that the elastic moduli of a 
diluted classical elastic lattice also obey scaling with the conductivity 
exponent, if the force constants are isotropic.\cite{degennes} With 
non-isotropic forces other scalings have been shown to be possible, and the 
critical dilution fraction above which the rigidity vanishes can in fact 
be below the percolation density.\cite{noncondexp} 

\begin{figure}
\centering
\epsfxsize=8.2cm
\leavevmode
\epsffile{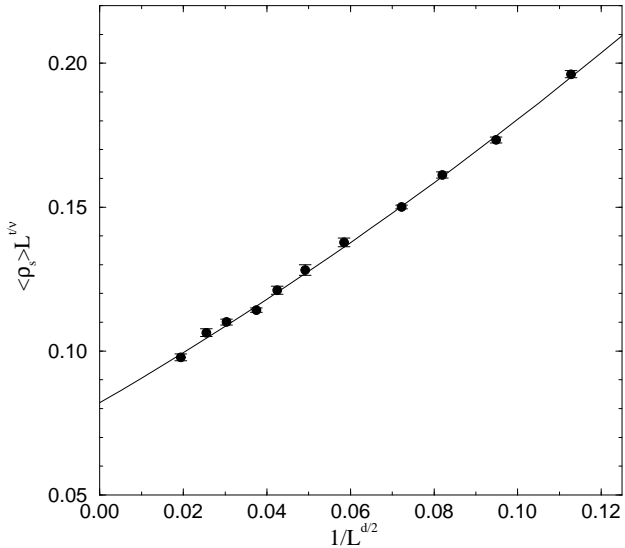}
\vskip1mm
\caption{Spin stiffness of site diluted $L\times L$ systems multiplied by 
$L^{t/\nu}$. The curve is a quadratic polynomial in $L^{-d/2}$.} 
\label{fig:rhol}
\end{figure}

The elastic moduli of classical percolating lattices have been studied 
extensively using numerical methods.\cite{noncondexp,lobb} The primary scaling
technique used there will be employed here as well. If the analogy with the 
random resistor network holds, the disorder averaged stiffness at the 
percolation point should scale as $L^{-t/\nu}$,\cite{lobb} where $\nu$ is the 
correlation length exponent of percolation, which is known exactly; $\nu=4/3$.
\cite{stauffer} Fig.~\ref{fig:rhol} shows the stiffness of the site-diluted 
Heisenberg model at the percolation density, multiplied by $L^{t/\nu}$, where 
$t/\nu = 0.9826$ was used.\cite{grassberger} The data extrapolate to a finite
value as $L\to \infty$, and hence the results are indeed consistent with the 
conductivity scaling. The average stiffness shown here was calculated by 
including only non-zero values of the stiffness in a given lattice direction, 
and was averaged also over the two equivalent directions. The simulations 
included only the largest cluster in the system. The fraction of non-zero 
stiffnesses is approximately $0.52$ for all $L$, in agreement with the known 
\cite{percdens} wrapping probability of clusters in periodic systems. 

\begin{figure}
\centering
\epsfxsize=8.2cm
\leavevmode
\epsffile{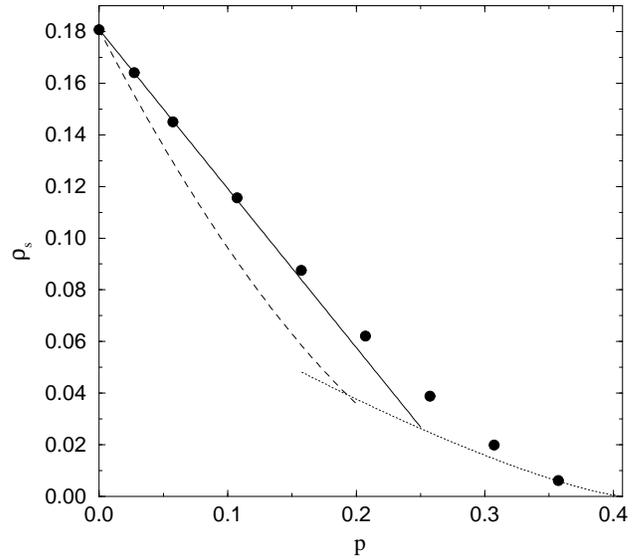}
\vskip1mm
\caption{Infinite-size extrapolated spin stiffness vs dilution fraction
(circles). The solid line is $\rho_s (p)=0.1808-0.62\cdot p$. The dotted
curve is the scaling form $\rho_s (p^*-p) \sim (p^*-p)^t$, and the dashed
curve is a non-linear $\sigma$-model result.\protect{\cite{castro}}} 
\label{fig:rhod}
\end{figure}

Fig.~\ref{fig:rhod} shows the full dilution dependence of the stiffness
extrapolated to infinite system size. The behavior is almost linear up to 
$p \approx 0.15$. The fitted linear form $\rho_s(p) = 0.1808 - 0.62\cdot p$ is
shown in the figure, along with an analytical result containing terms up to 
$p^2$ obtained using a non-linear $\sigma$-model and percolation theory in
Ref.~\onlinecite{castro}. The $\sigma$-model approach gives a quantum critical
point below the percolation point, and the initial fall-off is also 
faster than what is seen in the QMC data. The QMC data close to the 
percolation point are not well described by the random resistor network 
exponent, although the results at the percolation point indicate that this 
should be the asymptotic form as $p \to p^*$. The critical region may be 
very small, however, making it difficult to observe in direct calculations 
for $p < p^*$.

\section{Summary and discussion}

This paper has presented a variety of quantum Monte Carlo results showing that
the order-disorder transition in the diluted $S=1/2$ Heisenberg model is solely
driven by classical percolation. This is a consequence of the fractal clusters
at the percolation point having long-range antiferromagnetic order. The
presence of this long-range order was demonstrated by studying the largest 
cluster on $L\times L$ lattices, as well as clusters of fixed size $N_c$ on 
the infinite 2D lattice. For the infinite-size extrapolated sublattice 
magnetization, the same non-zero value was obtained for both types of 
clusters. An accurate calculation of the sublattice magnetization $M$ versus 
site dilution $p$ was made possible by taking advantage of a factoring into 
classical and quantum mechanical functions, which were evaluated separately. 
The classical factor is identical to the magnetization of a classical 
ferromagnet, and was calculated to high accuracy using lattice sizes $L$ up 
to $4096$. It contains the critical behavior of $M(p)$ close to the 
percolation point $p^*$. The quantum mechanical factor is equivalent to 
the sublattice magnetization of the largest cluster on $L\times L$ lattices 
in the limit $L \to \infty$. It remains non-singular as $p\to p^*$ and can be 
reliably extrapolated using relatively small system sizes (here using $L\le 
64$). Its infinite-size value grows only by a factor $2$ between $p=p^*$ 
and $p=0$. Approximate analytical forms describing the numerical $M(p)$ for 
all $p$ were also constructed. The spin stiffness at the percolation point
was shown to obey the same scaling as the conductivity of a random resistor 
network.

The conclusions reached in this paper differ from the non-universal quantum 
criticality scenario, which has been elucidated in several recent papers.
\cite{kato,todo,yasuda2} According to this scenario, the fractal clusters at 
the percolation point have power-law decaying spin-spin correlations, which 
implies that the scaling exponents differ from those of classical percolation.
It was argued that the exponents depend on the spin $S$ of the magnetic sites,
so that classical percolation is recovered only for $S\to \infty$.\cite{kato} 
Several types of scaling analyses have been presented in support of this 
unusual behavior.\cite{kato,todo,yasuda2}. It can be noted, however, that only
a very small number of system sizes were used. Temperature effects, although 
small,\cite{awscomment} may also have contributed to making the scaling appear
better than it would be for real $T=0$ data. 

The most serious problem with the finite-size scalings carried out in
Refs.\onlinecite{kato,todo}, and \onlinecite{yasuda2} is that even if the 
percolating clusters are ordered, as they in fact are, the staggered structure
factor cannot be expected to show the asymptotic classical scaling behavior 
for the range of system sizes used (as shown in Fig.~\ref{fig:spi}). 
This is due to the strong size dependence of the sublattice magnetization of 
the clusters (in contrast to the classical case, in which the cluster 
magnetization takes its maximum value at $T=0$ for any system size). The 
classical scaling form is valid only for systems sufficiently large for the 
relative size corrections to be small, which is the case only for systems much
larger than those that are currently accessible to quantum Monte Carlo 
simulation. This problem was circumvented here by focusing on the 
{\it cluster-size normalized} sublattice magnetization of the largest 
cluster of the lattice, which in combination with the known scaling of 
the cluster sizes completely determines the asymptotic behavior of 
the diluted system.

\begin{figure}
\centering
\epsfxsize=8.2cm
\leavevmode
\epsffile{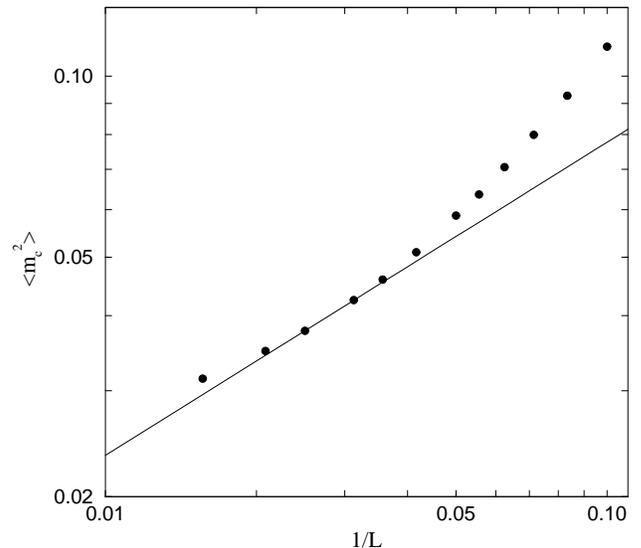}
\vskip1mm
\caption{The cluster magnetization vs inverse system size on a log-log
scale. The statistical errors are smaller than the circles. The line has 
slope $0.52$, which was previously argued \protect{\cite{todo}} to be the 
quantum critical scaling exponent.} 
\label{fig:logm}
\end{figure}

In Ref.~\onlinecite{todo} it was argued that the previous data 
\cite{awscomment} for the cluster magnetization for system sizes $L$ up to 
$48$ are also consistent with the quantum criticality scenario. On a log-log 
plot, the last few points were fitted to a straight line, and the same exponent
as that previously extracted from the staggered structure factor was 
obtained. The use of only a few system sizes in such 
a scaling is dangerous, however. It neglects the slow curvature that is 
evident in the data.\cite{todo} With one more system size now available 
($L=64$), as well as increased precision for smaller $L$, the failure of the 
quantum critical scaling can be demonstrated even more clearly. 
Fig.~\ref{fig:logm} shows a log-log plot of the cluster magnetization along 
with a line with slope $0.52$, which was previously argued to describe the 
data.\cite{todo} The $L=64$ point does not fall on the line, and also the 
smaller systems show deviations beyond the statistical errors. A slow upward 
curvature as $L\to \infty$ is evident. With no indication of 
a vanishing cluster magnetization on the linear scale in Fig.~\ref{fig:magln},
the log-log plot is clearly not suitable for analyzing the data.

In earlier Monte Carlo studies of disordered Heisenberg models
\cite{behre,awsmv} it was concluded that the antiferromagnetic order vanishes 
in a quantum phase transition before the classical percolation threshold. 
With the current results for much larger lattices at hand, it is clear that
state-of-the-art simulations at earlier times were not able to reach 
sufficiently large system sizes for observing the true asymptotic behavior. 
The essentially linear extrapolations used to extract the critical point were 
therefore misleading. Similar work on the disordered half-filled Hubbard model
on small lattices also have indicated that quantum fluctuations destroy the 
order before the percolation point.\cite{ulmke} In light of the behavior of 
the Heisenberg model, it would be interesting to repeat these calculations 
using larger lattice sizes. For the Hubbard model it is currently not 
possible to reach system sizes as large as for the Heisenberg model, however.

Experiments on quasi-2D cuprate antiferromagnets doped with non-magnetic
impurities have in the past been able to reach only dilution fractions 
$\alt 20\%$.\cite{doping} The doping dependence in this region is in 
reasonable agreement with calculations.\cite{castro} 
Recently, improved sample preparation techniques, involving simultaneous
doping with Zn and Mg, have enabled studies of La$\rm{_2}$CuO$\rm{_4}$ all 
the way to the percolation threshold.\cite{greven} Neutron scattering 
measurements of the temperature dependence of the correlation length and
the sublattice magnetization indicate that the order persists until 
$p \approx p^*$,\cite{greven} in accord with the behavior of the Heisenberg 
model discussed here. It can be noted that both the sublattice magnetization 
and the spin stiffness extracted in the experiments \cite{greven} agree
reasonably well with the curves extracted here (Figs.~\ref{fig:mfull} 
and \ref{fig:rhod}) at weak dilution but fall significantly faster
to zero as the percolation threshold is approached. This is an indication 
of additional quantum fluctuation mechanisms weakening (but not destroying) 
the long-range order, with likely candidates being frustrating 
next-nearest-neighbor interactions and 4-spin ring-exchange.\cite{coldea}
The effects of these interactions are likely more pronounced in the
diluted systems. Random lattice distortions causing fluctuations in the 
nearest-neighbor couplings could also play a role.

In a system exhibiting a quantum phase transition as a function of some
model parameter, e.g., the Heisenberg bilayer, \cite{bilayer} certain types
of dilution can drive an order-disorder transition before the classical
percolation threshold. If the dilution leads to magnetic moment formation 
the phase transition is destroyed, however, since the moments interact and 
order even in the gapped phase.\cite{wessel} In the bilayer, dilution of 
inter-layer dimers (two adjacent spins on opposite layers) does not lead to 
moment formation, and a quantum phase transition can occur before the 
percolation threshold (as a function of the dilution fraction or the 
inter-layer coupling). A multi-critical point where the percolating
cluster is quantum critical has recently been demonstrated in this system.
\cite{biperc} Quantum disordered ground states have also been found in 
Heisenberg antiferromagnets on non-random fractal lattices, such as the 
Sierpi\'nski gasket.\cite{gasket}

\acknowledgments

I would like to thank A. Castro Neto, A. Chernyshev, M. Greven, C. Henley, 
N. Kawashima, O. Sushkov, S. Todo, and N. Trivedi for discussions. This work 
was supported by the Academy of Finland (project 26175). Support from the 
V\"ais\"al\"a Foundation is also gratefully acknowledged. Some of the 
calculations were carried out using the Condor systems at the University 
of Wisconsin - Madison and the NCSA in Urbana, Illinois. 

\null\vskip-12mm\null


\begin{references}
\bibitem{memorynote}
In cases where memory size is a limiting factor, more efficient packing
of the data can further reduce the requirements. The size of the linked
list representing the configuration in the operator-loop update is also
strictly smaller than $M$, typically by about 20\%, since no $[0,0]$
elements need to be stored.
\end{references}
\end{document}